\def\USEACHEMSO{0} %

\if\USEACHEMSO1
\documentclass[journal=jctcce,manuscript=article,%
email=false,%
]{achemso}
\else
\documentclass[aip,jcp,notitlepage,twocolumn,
reprint,%
floatfix]{revtex4-1} %
\fi

\usepackage[utf8]{inputenc}
\usepackage[T1]{fontenc} 
\usepackage{stix}
\usepackage{mhchem}
\usepackage{hyperref}
\usepackage{units}

\usepackage{xspace}
\usepackage{verbatim}
\let\oldtheequation\theequation
\makeatletter
\def\tagform@#1{\maketag@@@{\ignorespaces#1\unskip\@@italiccorr}}
\renewcommand{\theequation}{(\oldtheequation)}
\makeatother 

\if\USEACHEMSO1
\fi

\usepackage{listings}

\usepackage[textsize=small]{todonotes}
\newcommand{\braket}[2]{\ensuremath{ \langle #1 | \, #2  \rangle }}
\newcommand{\Braket}[2]{\ensuremath{\left \langle #1 \vphantom{#1 #2}
    \left | \, #2 \vphantom{#1 #2} \right . \right \rangle }}

\newcommand{\ket}[1]{\ensuremath{  | {#1} \rangle}}

\newcommand{\matrixe}[3]{\ensuremath{ \langle{#1} | {#2} | {#3} \rangle }}

\newcommand{\erw}[1]{\ensuremath { %
    \langle {#1} \rangle}}

\newcommand{\dd}{\ensuremath{\mathrm{d}}}
\newcommand{\ii}{\ensuremath{\mathrm{i}}}
\newcommand{\cre}{\hat a^\dagger}
\newcommand{\ann}{\hat a}
\newcommand{\ie}{\mbox{i.\,e.\xspace}}

\newcommand{\Nel}{\ensuremath{N}}
\newcommand{\Na}{\ensuremath{N_\alpha}}
\newcommand{\Nb}{\ensuremath{N_\beta}}
\newcommand{\Ng}{\ensuremath{N_\text{grd}}}
\newcommand{\Norb}{\ensuremath{K}}
\newcommand{\bdim}{\ensuremath{D}}

\newcommand{\lit}[1]{Ref.~[\!\!\citenum{#1}]\xspace}

\definecolor{CBdblue}{RGB}{5,113,176}
\newcommand{\new}{}

\if\USEACHEMSO0
\begin{document}
\fi

\title{Minimal matrix product states and generalizations of mean-field and geminal wavefunctions}
\author{Henrik R.~Larsson}
\affiliation{Division of Chemistry and Chemical Engineering, California Institute of Technology, Pasadena, CA 91125, USA}
\author{Carlos A.~Jim\'{e}nez-Hoyos}
\affiliation{Department of Chemistry, Wesleyan University, Middletown, CT  06459, USA}
\author{Garnet Kin-Lic Chan}
\affiliation{Division of Chemistry and Chemical Engineering, California Institute of Technology, Pasadena, CA 91125, USA}

\if\USEACHEMSO1
\begin{document}
\fi

\begin{abstract}
Simple wavefunctions of low computational cost but which can achieve qualitative accuracy across the whole potential energy  
surface (PES) are of relevance to many areas of electronic structure as well as to applications to dynamics. Here, we explore a class of simple
wavefunctions, the minimal matrix product state (MMPS), that generalizes many  simple wavefunctions in common use,
such as projected mean-field wavefunctions, geminal wavefunctions, and generalized valence bond states.
By examining the performance of MMPSs for PESs %
of some prototypical systems, we find that they yield good qualitative behavior across the whole PES,
often significantly improving on the aforementioned ansätze.
\end{abstract}

\maketitle
\section{Introduction}

Simple qualitative wavefunctions, such as the Slater determinants used in Hartree-Fock (HF) and Kohn-Sham theory, play
essential roles in the theory of electronic structure.\cite{ringschuck_book,blaizotripka_book,jensen_book} %
For example, they provide qualitative understanding about bonding and structure,
and are a starting point for more sophisticated numerical treatments, 
via perturbation theory or as the dominant component in a more flexible ansatz.
In addition, because computations with such wavefunctions are cheap (often $N^3$ or $N^4$ cost where $N$ is proportional to system size)
such wavefunctions may be used both to study large systems, and to study dynamics, where cheap electronic structure methods are essential.

Beyond Slater determinants, other simple wavefunctions in common use 
can be thought of as small generalizations.
One class is obtained by breaking and restoring the symmetries in a Slater determinant.%
\cite{spinProjDMRG2017li,ringschuck_book,blaizotripka_book,projHFB2001scuseria,spinProj2019thompson,pHF2016scuseria,symmetryDilemma1963lykos,pEHFrev1980mayer,spinProjHF1955lowdin}
For example, typical Hamiltonians conserve particle number ($N$), spin symmetry ($S^2$, $S_z$), and time reversal symmetry, in addition to various point group symmetries. Rather than using a Slater determinant that obeys all these symmetries, one can break the symmetries in order
to capture essential correlations, and then restore them using projectors.
This leads to a variety of wavefunctions, such as projected unrestricted Hartree-Fock~\cite{spinProjHF1955lowdin,symmetryDilemma1963lykos,pEHFrev1980mayer} 
(broken and restored spin symmetry), the antisymmetrized geminal power
(AGP) (broken and restored number symmetry),\cite{coleman1965structure,hfbRev1975mang,ringschuck_book,blaizotripka_book}
and, as \new{an} extension to AGP, projected Hartree-Fock-Bogoliubov (HFB).\cite{ringschuck_book,projHFB2000sheikh,projHFB2001scuseria}
These wavefunctions are easy to compute with, because their mean-field origin
means that matrix elements can be obtained by a modified Wick's theorem.
Another way to create a simple wavefunction is to construct a product state not of orbitals,
but of multi-electron objects. The generalized valence bond (GVB) state is one such example, corresponding to a product state of strongly orthogonal
two-particle (geminal) wavefunctions.\cite{hurley1953molecular,gvb1967goddard,gvb1971goddard,gvbRev1973goddard,gvbRev2011hiberty,rassolov2002geminal,jensen_book,gvb1996cullen,gvbBeyond2017ayers,geminal2016limacher}

In this work, we describe another convenient way to generate simple wavefunctions using the formalism
of matrix product states (MPSs), the wavefunction ansatz of the density matrix renormalization group (DMRG).\cite{dmrgRev2011schollwock,dmrg1992white,dmrg1993white,dmrgRev2011chan,dmrgRev2020reiher,dmrgRev2015legeza}   
Matrix product states provide several ways to generalize the above pictures.
First, they allow for expectation values to be efficiently evaluated without the structure of a generalized Wick's theorem.
Second, it is natural to work with products of many-particle objects in the MPS form.
Third, by increasing the MPS bond dimension $\bdim$ (defined below) one can easily incorporate correlations beyond
those purely from symmetry projection, or contained within the individual wavefunction components (be they orbitals, geminals, or
more complex objects). Given the second quantized Hamiltonian, the
cost of a MPS calculation scales like $K^4$ (where $K$ is the number
of orbitals) with a prefactor that depends polynomially on the dimension  $\bdim$ of
the matrices that are the variational parameters of the state.\cite{qcDMRG1999white,dmrg2002chan,dmrg2004chan,dmrgRev2015legeza}
While in typical DMRG calculations, the bond dimension is made
very large in order to provide near exact answers, in the current work we focus on the opposite limit where $\bdim$ is very small, e.g. $\mathcal O(1)$,
and thus the prefactor in front of $\Norb^4$ is very small.
We shall call such states \emph{minimal} matrix product states (MMPS). As we shall see, in conjunction with symmetry projection, even the
smallest minimal matrix product state with $\bdim=1$ already encompasses the simple wavefunctions in common use, while generalizing to
new classes of simple wavefunctions that have not previously been considered. 

The remainder of this article is organized as follows: 
\autoref{sec:theory} gives an overview of the MMPS ansatz in (\autoref{sec:definitions}), its connection to geminal and related ansätze (\autoref{sec:theory_geminal}), and describes the algorithmic implementation of the MMPS ansatz (\autoref{sec:implementation}).
\autoref{sec:results} presents MMPS results for some prototypical systems and compares them to results from related ansätze. \autoref{sec:conclusions} concludes and gives our outlook on future applications.

\section{Theory}
\label{sec:theory}
\subsection{Minimal matrix product state ansatz}
\label{sec:definitions}

A matrix product state is obtained by writing the amplitude of a wavefunction as a product of matrices $\mathbf{A}^{n_i}$,
namely, for $\Norb$ orbitals
\begin{align}
  \ket{\Psi_\mathrm{MPS}} = \sum_{\{ n\}} \mathbf{A}^{n_1} \mathbf{A}^{n_2} \ldots \mathbf{A}^{n_{\Norb}} \ket{n_1 n_2 \ldots n_{\Norb}} \label{eq:mps_def}
\end{align}
where $\ket{n_1 n_2\ldots n_{\Norb}}$ is an occupancy vector for sites $1 \ldots \Norb$.\cite{dmrgRev2011chan,dmrgRev2011schollwock,dmrgRev2015legeza}
In the simplest case we consider,
we assume that the basis of site $i$ is a single orbital, i.e. $\{ \ket{n_i} \} = \{\ket{\mathrm{vac}}, \ket{\phi_i^\alpha}, \ket{\phi_i^\beta}, \ket{\phi_i^\alpha \phi_i^\beta}\}$. In a restricted formalism (used here) we further assume $\braket{r \alpha}{\phi_i^\alpha} = \braket{r\beta}{\phi_i^\beta}$.
The representational power of the MPS is controlled by the bond dimension of the matrices, which is $\bdim \times \bdim$ save for the
first and last which are $1\times \bdim$ and $\bdim\times 1$.

The smallest matrix product state is the simple product state with $\bdim=1$,
i.e. $\mathbf{A}^{n_i}$ is a scalar for each element of the site basis.
Such a state will not generally respect the symmetries of the system. Consequently, we
define a minimal matrix product state as the state obtained from the product state after an additional projection onto the pure
symmetry sectors of the Hamiltonian. In this work, we consider Hamiltonians where $N$, $S^2$ and $S_z$ are
good quantum numbers. Thus we define the  minimal matrix product state
to conserve one or more of these symmetries, e.g. 
\begin{align}
  |\Psi_\text{MMPS}\rangle = \hat P \ket{\Psi_\text{MPS}} =  \hat P^{S^2, S_z}  \hat P^N|\Psi_\text{MPS}\rangle \label{eq:mmps_def}
  \end{align}
where e.g.~$\hat P^N$ denotes projection onto a given particle number $\Nel$. Note that the distinction between MMPS and earlier
projected matrix product states such as the spin-projected MPS~\cite{spinProjDMRG2017li,li2019electronic} is mainly one of emphasis on using the smallest
bond dimensions.
While $|\Psi_\text{MMPS}\rangle$ is itself an MPS of a bond dimension given by that of the $|\Psi_\text{MPS}\rangle$ multiplied by that of the projector $\hat P$, the explicit larger representation never needs to be
formed in standard computations (see \autoref{sec:implementation} for more details).

It is useful to contrast the above scheme with how symmetries are usually expressed in MPSs without projection.\cite{dmrgRev2011schollwock,dmrg2002chan,dmrgSpin2012chan,dmrgSpin2016reiher,dmrgRev2015legeza}
For Abelian symmetries, such as $N$ and $S_z$, so long as $\{|n_i\rangle\}$ are eigenstates of $\hat N$ and $\hat S_z$, one can ensure \new{that} $|\Psi_\mathrm{MPS}\rangle$
is an irrep of these symmetries by requiring that the matrices $\mathbf{A}^{n_i}$ have a block structure.
Choosing reasonable sizes for such quantum number blocks is a discrete optimization process
that is challenging when the total bond dimension is small. In the projection approach, the need to choose a block structure is avoided,
which thus allows meaningful calculations with very small bond dimension, as small as $D=1$.

From the above definition of a MMPS, we can extend the ansatz in two natural ways. The first way is to enlarge the definition of a site in the
underlying MPS to capture the Hilbert space of multiple spin orbitals. 
For example, we may consider grouping pairs of
the above sites into single sites, e.g. $\{ |n_i n_{i+1} \rangle\new{\}} \to \new{\{}|\tilde{n}_{i/2}\rangle \new{\}}$, where the dimension of $\{\new{\ket{ \tilde{n}_{i/2}}}\}$
is now 16. The parent MPS is then still a product state, but of more complex components, similar to e.g.~a GVB state.
We shall refer to such MMPS as multisite MMPS. The second way is to increase
the bond dimension of $\mathbf{A}^{n_i}$ (i.e.~they become matrices) in the typical way that matrix product states are made more \new{accurate}.
As explained in the introduction, in this work we will focus on the case of small bond dimensions e.g.~$\bdim=1-5$, keeping the ansatz as minimal
as possible. In the evaluation of the computational costs (see \autoref{sec:implementation}), $\bdim$ thus enters only as a small prefactor.

It is important to note that, \new{similar to} normal MPS with insufficiently large $\bdim$,
the MMPS is not invariant to orbital transformations between sites (including the ordering of the sites).
Thus, as is the case for other simple wavefunctions,
its quality depends heavily on the orbitals used to define it. In numerical calculations, orbital optimization is thus often a necessary consideration. 

\subsection{Exponential form and connection to geminal powers and other ans\"atze} 
\label{sec:theory_geminal}

To more easily connect the {$\bdim=1$} MMPS to other commonly used simple wavefunctions, we first write it in
another explicit form. For the most direct correspondence, we first consider the case where the
sites are single orbitals. Then,
\begin{align}
|\Psi_\text{MMPS}\rangle = \hat P \mathcal{T} \prod_i (c_i + s_{i\alpha} \hat a^\dag_{i\alpha} + s_{i\beta} \hat a^\dag_{i\beta} + d_i \hat a^\dag_{i\alpha} \hat a^\dag_{i\beta}) |\mathrm{vac}\rangle \label{eq:mpsexplicitprod}
\end{align}
where the ordering operator $\mathcal{T}$ ensures that the non-commuting single creation operators are applied in lexicographical order
(note that the constants and double creation operators commute with each other and all single creation terms) e.g.
\begin{align}
  \mathcal{{T}} \prod_{i\sigma} \hat a^\dag_{i\sigma} = \hat a^\dag_{1\alpha} \hat a^\dag_{1\beta} \hat a^\dag_{2\alpha} \hat a^\dag_{2\beta} \ldots
  \end{align}
For the sites where $c_i \neq 0$, we can rewrite the factors in \autoref{eq:mpsexplicitprod} as exponentials 
 since $c e^{c^{-1}({s}_{\alpha} \hat a^\dag_\alpha + {s}_{\beta} \hat a^\dag_\beta + {d}\hat a^\dag_{\alpha} \hat a^\dag_\beta)} = c + s_{\alpha} \hat a^\dag_\alpha + s_{\beta} \hat a^\dag_\beta + d\hat a^\dag_{\alpha} \hat a^\dag_\beta$. Thus if all $c_i \neq 0$, the $D=1$ MMPS is an ordered exponential up to a scaling factor,
\begin{align}
  |\Psi_\text{MMPS}\rangle = \hat P   \mathcal{T} e^{\sum_{i\sigma} s_{i\sigma} \hat a^\dag_{i\sigma} + \sum_i d_i\hat a^\dag_{i\alpha} \hat a^\dag_{i\beta}}|\mathrm{vac}\rangle \label{eq:mpsexplicitprodexp}
\end{align}

The general AGP ansatz in its canonical basis (\ie, after an appropriate orbital rotation) with $N_s$ singly occupied orbitals can be written as %
\begin{align}
    \ket{\Psi_\text{AGP}} = \hat P^N \mathcal{T} \prod_{i=1}^{N_s} \cre_{i\alpha} \prod_{i={N_s+1}}^{\Norb} \left(1 + d_i \cre_{i\alpha} \cre_{i\beta}\right) \ket{\text{vac}} \label{eq:gagp}
\end{align}
Comparing this to the MMPS form of \autoref{eq:mpsexplicitprod} we see the MMPS reduces to the general AGP if for $N_s$ of the factors, we only
have one coefficient $s_{i\sigma}$ per factor, while for the other
factors, we only have the constant $c_i$ and double creation $d_i$ term,
reproducing the geminal terms in \autoref{eq:gagp}. Consequently, we refer to the latter factors as the geminal part
of the MMPS wavefunction.

Since the single site $D=1$ MMPS is distinguished from the AGP by the way in which the single creation
operators enter into the ansatz, we can compare also to some other wavefunctions which are related to the AGP but which introduce
single creation operators in a different way. Fukutome and coworkers introduced a generalization of the Bardeen-Cooper-Schrieffer wavefunction (the
AGP before projection) with single creation operators in an exponential,\cite{fermionLieAlg1977fukutome,fermionLieAlg1981fukutome} written as
\begin{align}
  |\Psi_\text{F}\rangle = e^{\sum_i (\theta_i \hat a^\dag_i -\theta_i^* \hat a_i)} e^{\sum_{i} d_{i} \hat a^\dag_{i\alpha}\hat a_{i\beta} } |\mathrm{vac}\rangle
  \label{eq:fukutome}
\end{align}
where $\theta_i$ are complex numbers. However, note that $ e^{\sum_i (\theta_i \hat a^\dag_i -\theta_i^* \hat a_i)} = c_0 + \sum_i (c_i \hat a^\dag_i - c_i^* \hat a_i)$ for
some constants $c_0, c_i$, thus this is very different from the MMPS where there is an ordered exponential; in particular, unlike
in the MMPS, if $d_i = 0$ it is not possible for the single creation operators to create a state with more than a single particle. Finally
we note that exponentials of single creation operators also occur in fermion coherent states similarly to in \autoref{eq:fukutome}, but there
$\theta_i, \theta_i^*$ are Grassman numbers.\cite{blaizotripka_book} This ensures that expectation values with fermion coherent states satisfy Wick's theorem for expectation values (i.e.~expectation values of fermionic operators can be expressed in terms of sums of products of single-particle density matrices)
but it also means that the amplitude of a fermionic coherent state is not physically meaningful, as it is a Grassman number.
        
To understand the variational freedom introduced by the single creation operators in the MMPS, we can consider a
simple limiting case where the geminal coefficients $d_i$ are 0 in
\autoref{eq:mpsexplicitprodexp}. This corresponds to
assuming all wavefunction amplitudes can be factorized as
\begin{align}
  \Braket{\phi_{i_1}^{\sigma_{i_1}} \phi_{i_2}^{\sigma_{i_2}}\ldots \phi_{i_N}^{\sigma_{i_N}} }{\Psi} = s_{i_1 \sigma_{i_1}} s_{i_2 \sigma_{i_2}} \dots
  s_{i_N \sigma_{i_N}}
\end{align}
The representational power of such a form is highly limited; it is not possible to doubly occupy any spatial orbital. There are nonetheless some non-trivial states that can be captured in this way.
In general, if we assume each $\alpha$ and $\beta$ orbital has the same spatial component, then
the single creation operators create an orbital of rotated spin (a generalized spin orbital), 
\begin{align}
  \sum_\sigma s_{i\sigma} \hat a^\dag_{i\sigma} |\mathrm{vac}\rangle  = \sqrt{\sum_\sigma |s_{i\sigma}|^2} |{\phi}_i^{\bar{\sigma}_i}\rangle
  \end{align}
where $\bar{\sigma}$ denotes the rotated spin. Incorporating projection onto fixed $N$, then the MMPS becomes  
a weighted distribution \new{over $N$-particle products} of generalized spin orbitals
\new{
\begin{equation}
\begin{split}
\ket{\Psi} = \sum_{i_1=1}^K& \sum_{\substack{i_2=2\\ i_2\neq i_1}}^K \sum_{\substack{i_3=3\\ i_3\notin \{i_1, i_2\}}}^K\cdots \sum_{\substack{i_N=N\\ i_N\notin \{i_1,i_2,\cdots,i_{N-1}\}}}^K \\%
&c_{i_1}c_{i_2}\cdots c_{i_N} %
 \ket{\phi_{i_1}^{\bar{\sigma}_{i_1}} \phi_{i_2}^{\bar{\sigma}_{i_2}} \ldots \phi_{i_N}^{\bar{\sigma}_{i_N}}}
 \end{split}
\end{equation}
}
where $c_{i} = \sqrt{\sum_\sigma |s_{i\sigma}|^2}$. 
\new{
For any $K>N$, this represents a non-trivial linear combination; for example, for $K=3$ and $N=2$, we get $\ket{\Psi} = c_1 c_2 \ket{\phi_{1}^{\bar{\sigma}_{1}}\phi_{2}^{\bar{\sigma}_{2}}} + c_1 c_3 \ket{\phi_{1}^{\bar{\sigma}_{1}}\phi_{3}^{\bar{\sigma}_{3}}} + c_2 c_3\ket{\phi_{2}^{\bar{\sigma}_{2}}\phi_{3}^{\bar{\sigma}_{3}}} $. 
Thus even this \emph{artifically simple} ($d_i=0$) example of an MMPS describes physics different than that of other mean-field and projected mean-field states.
}

As another example, note that an AGP state is written as a linear
combination of all doubly occupied determinants but the AGP ansatz
does not include determinants from higher seniority sectors.
In the MMPS, the inclusion of the single creation operators via the
ordering operator $\mathcal{T}$ yields a state that can formally access all determinants in the Hilbert space.

Multisite MMPS, as well as bond dimensions with $\bdim>1$ have the potential to compactly represent even more qualitative electronic structure\new{s} beyond
that captured by the AGP language. For example, the perfect pairing GVB wavefunction~\cite{jensen_book,gvbRev2011hiberty}
can be written (up to normalization) as
 \begin{align}
   |\Psi_{\mathrm{GVB}}\rangle = \prod_{i=1}^{\Norb/2} (\hat a^\dag_{i\alpha} \hat a^\dag_{i\beta} + d_i \hat a^\dag_{\bar{i}\alpha} \hat a^\dag_{\bar{i}\beta}) |\mathrm{vac}\rangle
 \end{align}
where indices $i, \bar{i}$ index the perfect pairing orbitals.
As this is a product state, it is clearly a matrix product state, and if the MPS sites are chosen to consist of the paired orbitals $\{\phi_{i\sigma},\phi_{\bar{i}\sigma'}\}$ then it is a MPS (and thus MMPS) of bond dimension 1. However, it is easy to generalize the perfect pairing GVB wavefunction now also
to include broken pairs by including the linear terms in the MMPS ansatz, or to include broken and restored symmetries, or to include
clusters of larger sites. The key point is that formulating the ansatz in the matrix product language provides a simple organization of the computation,
which does not require the unprojected state to obey Wick's theorem for expectation values (as for projected mean-field and AGP states) or to be a
single product state (as for GVB).

\new{As with many of the other wavefunction ansätze discussed, MMPS (and MPS) are not size consistent in general. For normal MPS, size
consistency requires an appropriate choice of orbitals and their ordering. For the MMPS, size consistency is broken by the projector but
recovered (for the appropriate choice of orbitals and ordering) in the large $D$ limit. Nonetheless, in many cases of interest the extensive scaling
of the correlation energy is less important than the treatment of the intensive changes in the correlation energy in a local region where bonds are
changing, which the MMPS can recover using orbitals localized to that region.
In addition, by imposing local particle number constraints on the projector, global size consistency can be restored, as has been demonstrated with the  Jastrow-AGP ansatz.\cite{sizeConstAGP2020neuscamman} However, this is beyond the scope of this work.
}

\subsection{Implementation}
\label{sec:implementation}

The variationally minimized energy of the MMPS ansatz~\autoref{eq:mmps_def} can be carried out using the following functional~\cite{ringschuck_book}
\begin{equation}
 E = \text{min}_{\Psi_\text{MPS}} \frac{\matrixe{\Psi_\text{MPS}}{\hat H\hat P}{\Psi_\text{MPS}}}{\matrixe{\Psi_\text{MPS}}{\hat P}{\Psi_\text{MPS}}},\label{eq:tise_p}
\end{equation}
where we have used the fact that $\hat P$ commutes with $\hat H$ and idempotency of $\hat P$. Note that $\ket{\Psi_\text{MMPS}}$ does not explicitly appear in \autoref{eq:tise_p} and thus does not need to be constructed.
In the following, we describe possible numerical choices of $\hat P$ and the implementation of \autoref{eq:tise_p}.

\subsubsection{Choice of projector}

\label{sec:projector}
There are many ways to evaluate the expectation value of a projected wavefunction occurring in \autoref{eq:tise_p}. 
For example, in variational Monte Carlo, one samples the wavefunction using states that have the desired symmetries.\cite{sandroagp,sandroagp2,vmcMps2007sandvik,neuscamman2012optimizing,mahajan2019symmetry,spinfciqmc2019dobrautz} 
Here, we use an explicit operator
representation of the projector. Formally, a  projector is a delta distribution
that selects the eigenstates to project on.\cite{ringschuck_book,projectors2007izmaylov}
For example, $\hat P^N = \delta(\hat{N} - N)$ where $\hat N=\sum_i \hat n_i=\sum_i \cre_i\ann_i$.
We consider two explicit constructions of the projector:  a matrix-product-operator~\cite{dmrgRev2011schollwock} (MPO) construction, and an integral-based construction.

\begin{figure}[!tbp]
  \includegraphics[width=\columnwidth]{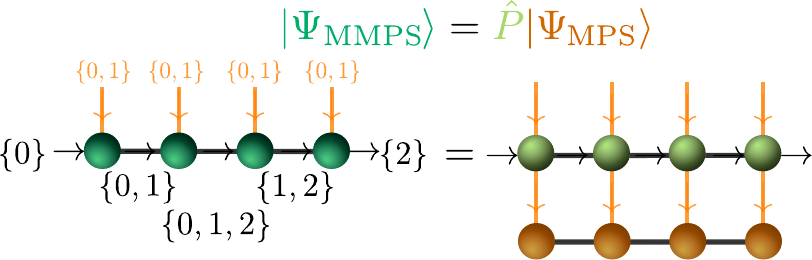}
  \caption{Diagram of the matrix-product-operator (MPO) based projector $P$ for constraining particle-number symmetry in the minimal matrix product state (MMPS). Applying $\hat P$ to the MPS (right-hand side) generates a MMPS (left-hand side). The MMPS formally corresponds to a MPS whose matrices are in
    block-sparse form with equally sized blocks, where each block is labelled by the combinations of particle numbers along the arrows. The arrows denote the ``flow'' of particles. The sets denote the particle number sectors on a particular bond in the MPS. An example is shown for $\Nel=2$ and four spin orbitals/sites.}
  \label{fig:projector_mpo}
\end{figure}

To illustrate the idea behind the MPO construction, we consider the representation of $\hat P^N$.
We define our MPO projector such that applying $\hat P^N$ to the MPS formally yields an MMPS
with the same structure as ordinary MPS with quantum numbers, i.e.~the
matrices $A^{n_i}$  have a block-structure labelled by particle number.
A pictorial example is shown in \autoref{fig:projector_mpo}.

To obtain the projector, we first, as in conventional DMRG,\cite{exStateGeomOptDMRG2015hu,dmrgRev2015legeza} %
construct all possible particle sectors for a given bond such that the initial site starts with $0$ and the last site ends with $\Nel$ electrons (compare with \autoref{fig:projector_mpo} for $\Nel=2$).
The number of particle sectors  on the bond corresponds then to the dimension of the MPO on that bond. The MPO tensor on a site $i$ is a matrix
for each bra, ket pair $n_i, n_i'$  in the basis of site $i$, and
to satisfy particle number balance the elements of the tensor take the form
\begin{equation}
  [\mathbf{M}^{n_i,n_i'}]_{lr} = \delta_{n_in_i'} \delta_{ N(l) + N(n_i), N(r)},\label{eq:pmpo_def}
\end{equation}
where $N(n_i)$ is the number of particles in state $|n_i\rangle$ i.e, $\{0,1\}$ for one spin orbital, and $N(l)$ and $N(r)$ are the number
of particles associated with the left bond index $l$ and right bond index $r$ of the MPO tensor.

For $\hat P^N$, the maximal bond dimension (maximal number of particle sectors at a given bond) is $D_{P}^{(N)} = N+1$ (this is the total
number of partitions of $N$ between the left and right halves of the system, i.e.~$(0, N), (1, N-1), \ldots (N,0)$).
Generalizing to a projector that fixes both $S_z$ and $N$, $\hat P^{\Na}\hat P^{\Nb}=\hat P^{S_z}\hat P^N$, the maximal bond dimension becomes
$D_P^{(\Na,\Nb)}=${$(\Na+1)(\Nb+1)$}. The projector can be generalized to $S^2$ symmetry by defining the tensor elements in
 \autoref{eq:pmpo_def} in terms of the Wigner $3j$ symbols.
 The MPO projector form has the property that the symmetry is directly encoded in the block structure of the MPO.
 However, while it works well in its exact form, we have found that it is not so easy to approximate at lower cost, as
 ``pruning'' the projector does not preserve the commutation betwen $\hat H$ and $\hat P$, required for the variational bound
 on the energy functional in \autoref{eq:tise_p}.

Alternatively, the projector can be constructed in an integral representation.\cite{ringschuck_book,hfbRev1975mang} For $\hat P^N$,
this takes the form
\begin{align}
  \hat P^N &= \frac{1}{2\pi} \int_0^{2\pi} \exp[\ii \phi (\hat N-\Nel)] \dd \phi.
\end{align}
Discretizing the integral with $\Ng$ grid points gives
\begin{align}
  \hat P^N &= \frac{1}{\Ng} \sum_{n=0}^{\Ng-1} \exp[\ii \phi_n (\hat N - \Nel)] = \delta_{\hat N,\Nel}, \label{eq:discrete}
\end{align}
with $\phi_n = 2\pi n /{\Ng}$. Since $\exp(\ii \phi \hat N) = \exp(\ii  \phi \sum_i \hat n_i) = \prod_i \exp(\ii  \phi\hat n_i)$, %
$\hat P^N$ can be written as sum of products i.e., a sum of MPOs with $D=1$, or a single sparse MPO with bond dimension $\Ng$.\cite{spinProjDMRG2017li} This allows for an embarrassingly parallel implementation. 
While the overall $\hat P^N$ is real-valued, the individual terms $\exp(\ii \phi \hat N)$ in \autoref{eq:discrete} are complex-valued.
However, to avoid complex algebra, \autoref{eq:discrete} can be recast into a sum over $\Ng/2$, $D=2$ real-valued MPOs, plus one $D=1$ MPO term (or a single block-sparse MPO with bond dimension $\Ng+1$) by 
using only the real-valued part of the individual terms:
\begin{align}
 \hat P^N &=  \frac{1}{\Ng} \sum_{n=0}^{\Ng-1} \cos[\phi_n (\hat N - \Nel)]\\
  &= \frac{1}{\Ng} \hat 1 +\frac{2}{\Ng} \sum_{n=1}^{\Ng/2} \cos[\phi_n (\hat N - \Nel)] \label{eq:Pn_rsum}
\end{align}
where we made use of %
the periodicity and the even symmetry of the $\cos$ function
and assumed odd $\Ng$.\footnote{For even $\Ng$, the last sum-term in \autoref{eq:Pn_rsum} has to be multiplied by $1/2$.}
$\cos[\phi_n (\hat N - \Nel)]$ can then be written as an MPO via
 \begin{align}
   \begin{bmatrix} c(\hat n_1' \phi_n) & s(\hat n_1' \phi_n)\end{bmatrix}
   \begin{bmatrix} c(\hat n_2' \phi_n) & s(\hat n_2' \phi_n) \\
       -s(\hat n_2' \phi_n) & c(\hat n_2' \phi_n)\end{bmatrix}
   \cdots
     \begin{bmatrix} c(\hat n_K' \phi_n) \\  -s(\hat n_K' \phi_n)\end{bmatrix}
 \end{align}
where we have used the shorthand $c(\phi) = \cos \phi$, $s(\phi) = \sin(\phi)$ and $\hat n_i' = \hat n_i - N / \Norb$.

While we found numerically, for the cases we have studied (up to $\Norb=8$ and $\Nel=8$),
that the complex-valued sum can also be fitted into a real-valued sum (of $D=1$ MPOs) with twice as many terms, the numerical fitting procedure~\cite{tensorDecompRev2009kolda} is difficult and not well-conditioned if $\Norb$ is \new{large}. Thus, we
leave the question of other simple, analytical real-valued descriptions of \autoref{eq:discrete} for future considerations.  Instead, in the following we
stick to the slightly more computationally demanding real-valued $D=2$-MPO-form, which has off-diagonal terms in each MPO.

Similarly to $\hat P^\Nel$, a projector onto fixed $S_z$ and $S^2$ can also be constructed in integral form~\cite{ringschuck_book,spinProjDMRG2017li,projHFB2001scuseria}
\begin{align}
  \hat P^{S^2,S_z} &= \hat P^{S_z} \hat{\mathcal P}^{S^2} \hat P^{S_z},\label{eq:P_SzS2}\\
  \hat{\mathcal P}^{S^2} &= \frac{2S+1}{2} \int_0^\pi  \sin(\alpha) d_{M,M}^{S}(\alpha) \exp(-\ii \alpha \hat S_y)\dd \alpha,
\end{align}
where $d_{M,M}^{S}$ is the small Wigner-D matrix.\cite{joachain_book}
$\hat{\mathcal P}^{S^2}$ is not a true projector and mixes spin orbitals, thus $\hat P^{S_z}$ has to be applied twice in \autoref{eq:P_SzS2} in order to ensure that $\hat P^{S^2,S_z}$ is a projector.
$\hat P^{S_z}$ is  defined analogously to $\hat P^N$ and can be implemented in the same manner. 
$\hat P^{S^2}$ can be evaluated via  Gauß-Legendre quadrature and results in a real-valued sum of terms. The number of
quadrature points required to evaluate the integral exactly is stated in \lit{spinProjDMRG2017li} and
is proportional to the number of singly occupied orbitals and the $S$ value.

One advantage of the integral based construction is that one can easily obtain approximate projectors of lower cost by reducing the
number of grid points $\Ng$ in the integration. Although the approximate projectors no longer commute with $\hat H$ exactly, we have
found this to be less of an issue in practice than for the MPO based projector. 
We note that sufficient grid points have to be chosen for $\hat P^{S_z}$ in order to achieve idempotency of $\hat P^{S^2,S_z}$. In contrast, regardless of the number of grid points, $\hat P^{S_z}$ and $\hat P^N$ are always idempotent as they project onto $S_z$ or $N$ modulo $\Ng$.

\subsubsection{DMRG algorithm}
The standard way to optimize the energy of an MPS \emph{ansatz} is the DMRG algorithm, where, similar to an alternating least squares algorithm,\cite{tensorDecompRev2009kolda}
one optimizes a small number of (neighboring) sites $\{\mathbf{A}^{n_i},\mathbf{A}^{n_{i+1}},\dots, \mathbf{A}^{n_{i+x}}\}$ at a time while fixing the remaining sites $\{\mathbf{A}^{n_1},\dots, \mathbf{A}^{n_{i-1}}, \mathbf{A}^{n_{i+x+1}},\dots, \mathbf{A}^{n_{\Norb}}\}$.\cite{dmrg1992white,dmrg1993white}
This local optimization problem is quadratic and can be solved as an eigenvalue problem.
After some sites are optimized, the next neighboring sites are chosen until all sites in \new{the} MPS have been optimized. This is called a sweep and repeated until convergence.

For quantum-chemical Hamiltonians, the DMRG algorithm can be efficiently implemented using complementary operators.\cite{complOp1996xiang,qcDMRG1999white,dmrg2004chan}
The complementary operators consist of a precontraction of some of the terms in the Hamiltonian
which  provide an optimal way to use the sparsity existing in the Hamiltonian's MPO representation.\cite{dmrgMPO2015keller,mpo2016chan}
Here, to optimize the energy functional in \autoref{eq:tise_p} for the MMPS, we implemented a new DMRG code.
Specifically, we use a generalized implementation   that evaluates
\autoref{eq:tise_p} using the combined operator $\hat H \times \hat P$ within the complementary operator approach.
$\hat P$ is constructed using either the MPO or integral based construction as described in \autoref{sec:projector}
and is a sparse MPO of bond dimension $\bdim_P$. Because of the sparsity of the representation of $\hat P$, the MPO tensors have only $\mathcal{O}(\bdim_P)$
non-zero entries. 

Compared to a conventional DMRG implementation without $\hat P$, for each complementary operator of the Hamiltonian on a given site, there are
$\bdim_P$ associated terms to be stored. (The number of terms is proportional to $\bdim_P$ rather than $\bdim_P^2$ due to the MPO sparsity).
Further, the \emph{individual} terms in $[\mathbf{PH}]$ are non-symmetric as, e.g. $\cre_i \hat P \neq \hat P \cre_i$.
Hence, compared to a normal DMRG implementation, $2\bdim_P$ more terms need to be computed.
Note, however, that the formal bond dimension of the MMPS, obtained by applying $\hat P$ to the underlying MPS of bond dimension $D$,
is $D \times D_P$, and the cost of optimizing the MMPS is much cheaper than the cost of a DMRG computation with a general MPS of bond dimension $D \times D_P$. 
\new{The reduced cost can be understood in terms of the smaller number of parameters to be optimized (smaller matrices to be diagonalized) and by the simple form of $\hat P$. Whereas in conventional DMRG, all $D_P$ blocks of size $D\times D$ contain different values in the block-sparse MPS, in the MMPS, the blocks are all generated via $\hat P$ from a single block.}
Essentially,  introducing $\hat P$ shifts some computational effort from the MPS to the operator, at the cost of some restriction
in the degrees of freedom.

To allow for multisite MMPSs, we generalized the code to include an arbitrary selection of determinants on a given site.  
For $D=1$ this also enables AGP, GVB and similar wavefunction optimization, while for $D>1$ one can
optimize in the subspace of determinants included in the AGP or GVB ansätze.

With the aforementioned modifications, the remainder of the optimization can follow the normal DMRG algorithm.
Here, we used the one-site algorithm, where just one site is optimized at a time, in combination with perturbative noise to avoid getting stuck in local minima.\cite{dmrg1s2005white} 
For optimizing a particular site $i$, a generalized eigenvalue problem results from \autoref{eq:tise_p}:
\begin{equation}
[\mathbf{HP}] \mathbf{A}^{n_{i}} = \mathbf{P} \mathbf{A^{n_i}} E.
\end{equation}
Due to the  null space of $\hat P$,  the matrices  $[\mathbf{HP}]$ and $\mathbf{P}$ are indefinite and share the \emph{same} null space.
For some methods, this null space needs to be projected out.\cite{pCCnullSpace2019tsuchimochi}
Here, this costly projection can be avoided by using the Davidson method~\cite{davidson_1975} for generalized eigenvalue problems and by using an initial trial solution $\tilde{\mathbf{A}}^{n_s}$ that \new{is an element of} the kernel of $\hat P$. 
Only for poor approximations of $\hat P$ with insufficient quadrature points did we find numerical issues due to the null space of $\hat P$.

In most situations, even for a $D=1$ MMPS with only four parameters per site, the one-site DMRG algorithm performed well as an optimization
algorithm in our studies.
However, especially when non-optimal orbitals were used for the MMPS, a gradient-based optimization of the MMPS parameters instead of a DMRG optimization turned out to be more efficient in some cases. In practice, for these difficult cases, we used a combination of both DMRG and gradient-based trust-region methods.\cite{wright_optimization_book} 
For difficult cases such as the \ce{H4} system with AGP orbitals, we also performed basin hopping to avoid getting stuck in high-lying local minima.\cite{basinHopping1997wales}

Orbital optimization was performed using the \textsc{PySCF} quantum chemistry package,\cite{pyscf,pyscfnew} 
which requires the one- and two-body density matrices as input.\cite{casscf2017sun} These we computed as expectation values of the MMPS wavefunction.

\subsubsection{Computational cost}
\label{sec:impl_comp_cost}
Following the scaling analysis of the standard quantum-chemistry DMRG algorithm,\cite{dmrg2002chan,dmrg2004chan,dmrgRev2015legeza}
the computational cost of evaluating and optimizing the MMPS energy in \autoref{eq:tise_p} (given the second quantized integrals)
scales as $\mathcal O[C (\Norb^3 D^3 +\Norb^4 D^2)]$, where $C$ is the cost of applying the projector.
Here, the bond dimension $D$ is of $\mathcal O(1)$ so we write the cost more succinctly as $\mathcal O(C \Norb^4)$.
This scaling is the same as that of projected HFB, AGP, and other related methods.
If orbital optimization is performed, there is an additional 
$\Norb^5$ cost from \new{the} integral transformation \new{in} each orbital optimization step.

Because the projector is sparse in both the MPO and integral construction, $C$ is directly proportional
to the projector bond dimension $\bdim_P$. Thus for exact projectors, $C$ depends on the number
of symmetries projected against. For example, if we use $\hat P = \hat P^{S_z} \hat P^N$, then for the MPO construction
$C \propto \Na \Nb$, while for the integral form $C \propto \Ng = \Norb^2$. As mentioned above, we observe that approximate projectors
constructed in the integral form by using a reduced number of grid points $\Ng$ in practice work quite well.
Indeed, for mean-field-like methods such as HFB it has been observed that the required $\Ng$ scales better than linearly with system size for $\hat P^{N}$.\cite{projHFB2001scuseria} 
Also sparse cubature can reduce $\Ng$ for spin projection in HF.\cite{pGHF2018li}
However, we are not aware of rigorous studies of the scaling of the approximation error with system size due to a reduced $\Ng$, and we leave this question for future considerations.
\section{Results}
\label{sec:results}

We now study the behavior of the MMPS and the multisite MMPS  (i.e.~where a single site spans multiple orbitals) for some
prototypical problems that exhibit static correlation, and compare to results from similar ans\"atze such as GVB and AGP.
The systems we study are the \ce{H4} ring (\autoref{sec:res_h4}), \ce{O2} dissociation (\autoref{sec:res_o2}), and \ce{HF} dissociation (\autoref{sec:res_fh}). 

If not mentioned otherwise, the projector used for the MMPS is $\hat P=\hat P^{S_z}\hat P^{N}$.  For this projector, we use the MPO form 
defined in \autoref{sec:projector}. We will also use $\hat P=\hat P^{S^2, S_z} \hat P^N$. In this case, we employ the integral form defined in \autoref{sec:projector}
with $\Ng = 5$ grid points for the $N$ and $S_z$ integrations, and $\Ng=2$ grid points for the $S^2$ integration.
Unless stated otherwise, orbitals in the MMPS calculations were ordered according to canonical order (energy order
for HF orbitals, natural orbital occupancy for AGP orbitals, and in the same order as the starting
orbitals when using optimized orbitals).

Both MMPS and GVB optimization used the code described in \autoref{sec:implementation}. 
AGP optimization (except restricted open-shell (RO)-AGP) used code developed by one of the authors (CAJH). %
{Unless stated otherwise, we refer to restricted AGP when we use the term AGP and will explicitly state when we use unrestricted (U)-AGP.}
\subsection{\ce{H4} ring}
\label{sec:res_h4}
\begin{figure}[!tbp]
  \includegraphics[width=.6\columnwidth]{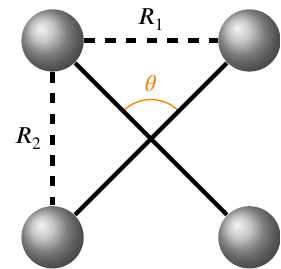}
  \caption{Geometry of the \ce{H4} ring. {The gray spheres denote the hydrogen atoms. $\theta$ denotes the angle to be scanned, which changes the bond distances $R_1$ and $R_2$ simultaneously.}}
  \label{fig:H4_setup}
\end{figure}

The \ce{H2 + H2} system is a prototypical system that at certain geometries exhibits strong multireference character.%
\cite{h4I1969goddard,h4II1972goddard,uhf1975fukutome,pHFpCC2017scuseria,ccH41998kowalski,mrccH42008evangelista} 
In the following, we place \ce{H4} on a ring of radius $\unit[3.3]{a_0}$ and scan the bond angle $\theta$ to obtain a potential energy curve (PEC; see \autoref{fig:H4_setup}).\cite{pHFpCC2017scuseria} The bond distances $R_1$ and $R_2$ are equal at the transition state (TS; $\theta=90^\circ$), %
and the ground and the first excited states are nearly degenerate when using a minimal basis STO-3G~\cite{sto3g}(as used here).

\begin{figure}[!tbp]
  \includegraphics[width=\columnwidth]{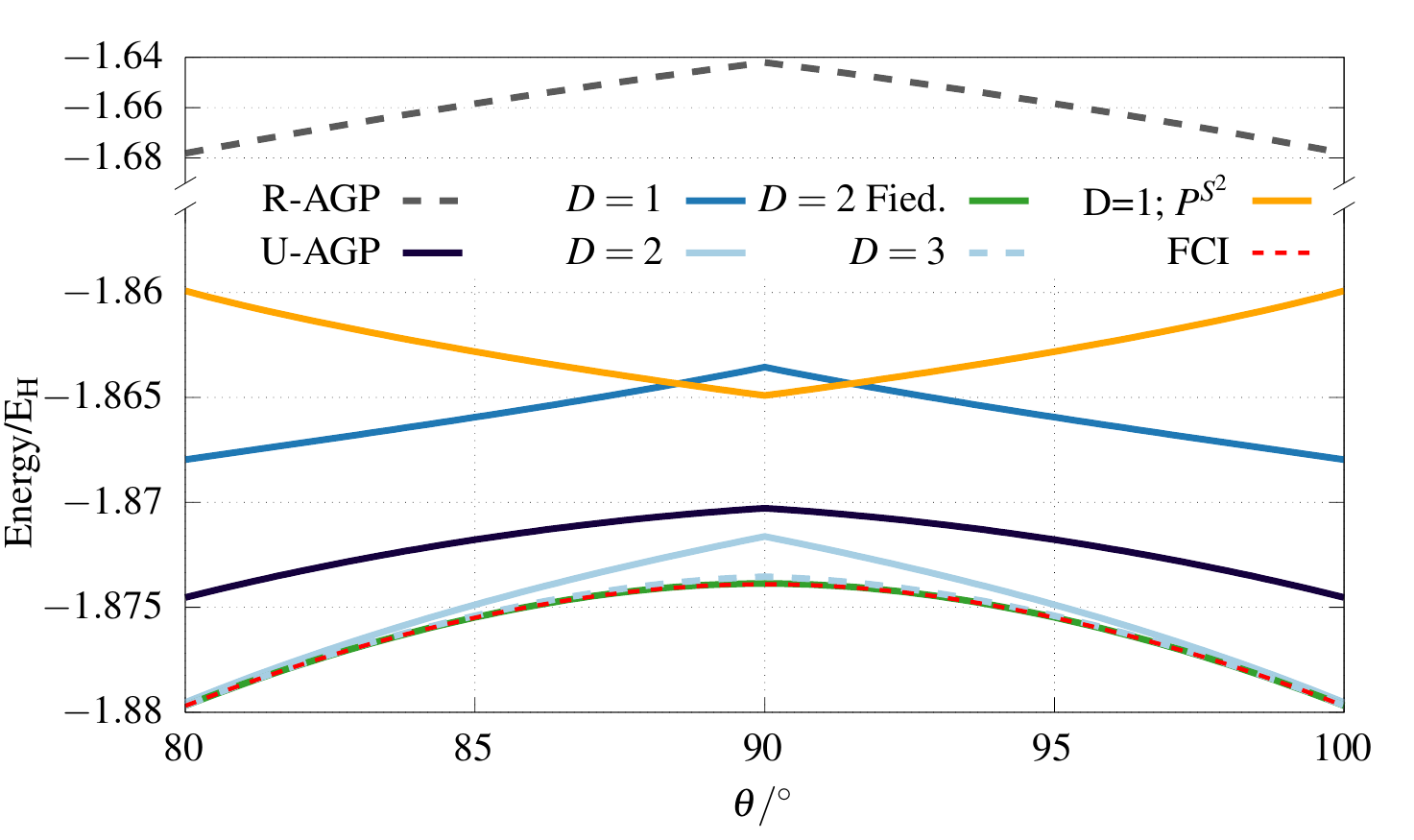}
  \caption{Potential energy scan for the \ce{H4} ring as depicted in \autoref{fig:H4_setup}. 
  {Shown are the results for (un-)restricted antisymmetrized geminal product, R(U)-AGP in dashed gray (black), minimal matrix product state (MMPS) with different bond dimension, $D$,  (blue, green, and orange) in comparison 
  to the full configuration interaction reference (FCI; dashed red curve).
  The MMPS curves use the R-AGP natural orbitals and, as a projector, $\hat P=\hat P^{S_z}\hat P^{N}$ (blue and green), and $\hat P=\hat P^{S^2, S_z} \hat P^{N}$ (orange), respectively.
 The green curve (on top of the FCI curve) denotes the MMPS $D=2$ result with R-AGP orbitals but ordered according the Fiedler vector at $\theta=80^\circ$.}
  The STO-3G basis is used.}
  \label{fig:H4_pes_1}
\end{figure}

The MMPS energies (blue curves) using restricted AGP (R-AGP) natural orbitals are shown in \autoref{fig:H4_pes_1} and compared to restricted and unrestricted AGP (U-AGP) (dashed gray and black curves). Compared to R-AGP, the $D=1$ MMPS already significantly improves the energy, both in an absolute sense
and in terms of \new{non-}parallelity to FCI.  Increasing the bond dimension slightly, we find that the $D=2$ MMPS yields lower energies even than U-AGP.
The curve retains an artificial cusp at $\theta=90^\circ$, but the $D=3$ MMPS PEC is
smooth and approximates the full configuration interaction (FCI; dashed red curve) result very well.

Due to the near degeneracy of excited states in this system with different spin, spin contamination is an issue for approximate methods.
The $D=1$ MMPS actually describes the first excited (triplet) state. In fact, when R-AGP orbitals are used,
the lowest stable singlet solution within the $D=1$ MMPS form corresponds to the R-AGP state.%
\footnote{Note that when using the same orbitals, the AGP state  always is a local extremum in the $D=1$ MMPS ansatz.}
While the $D=3$ curve reproduces the PEC well,  
spin contamination is still sizable and at the TS the $D=3$ MMPS has $\erw{\hat S^2}=0.1$.
Including $\hat P^{S^2}$ in the projector for the $D=1$ MMPS ensures that we find a singlet state (orange curve), but when using R-AGP orbitals this leads to 
qualitatively wrong energetics with a minimum at the actual TS.
\footnote{This can partly be attributed to the poor character of the R-AGP orbitals which break the symmetry of the nuclear framework.}

{We also performed MMPS calculations ordering the orbitals according to the Fiedler vector of the exchange matrix as is commonly
  performed in standard DMRG calculations~\cite{dmrgQI2011barcza,dmrgRev2015chan} at $\theta=80^\circ$.
For $D=1$ (not shown), AGP natural orbital ordering is better and Fiedler ordering leads to an MMPS with increased energy of $\sim\!8\cdot \unit[10^{-4}]{E_H}$.
However, for $D=2$ (green curve) Fiedler ordering greatly improves the energies and, already for $D=2$, they have an absolute error of only $\sim\!\unit[10^{-5}]{E_H}$, compared to the FCI energies.
}

\begin{figure}[!tbp]
  \includegraphics[width=\columnwidth]{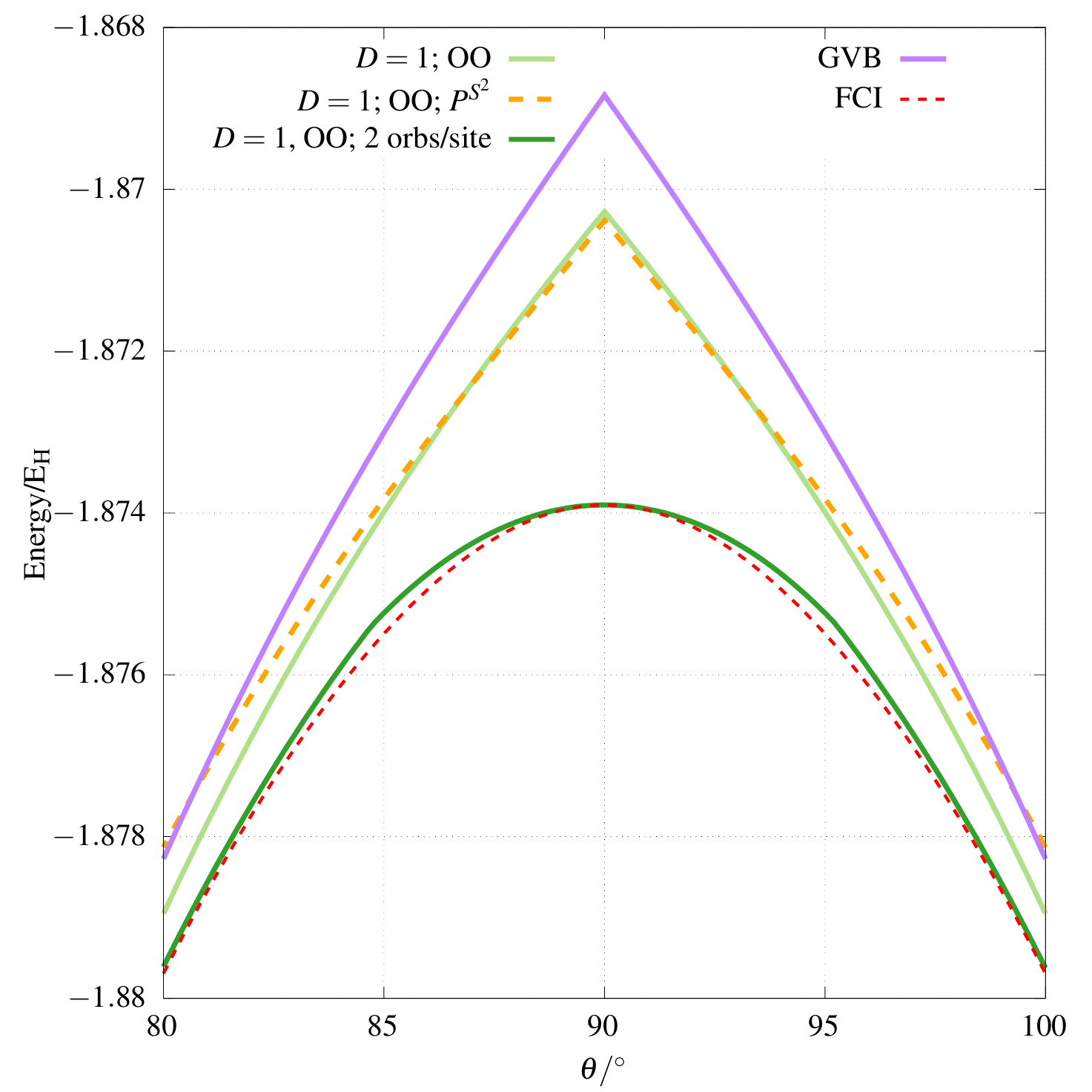}
  \caption{Same as \autoref{fig:H4_pes_1} but with orbital optimization {(OO)} for the {minimal matrix product state (MMPS)} and comparing to {generalized valence bond (GVB; purple) results.}
  The dark green curve denotes a multisite MMPS consisting of two spatial orbitals, a similar grouping to that used in the GVB state.
  }
  \label{fig:H4_pes_2}
\end{figure}

Besides orbital ordering, orbital optimization greatly improves all the MMPS results (\autoref{fig:H4_pes_2}), including for $D=1$.%
\footnote{Additionally, with orbital optimization, the MMPS state optimization is easier as there seem to be fewer high-lying local minima across the MPS parameter landscape, compared to when using non-optimal orbitals.}
Thus, when orbital optimization is included, the PEC of the $D=1$ state (pale green curve) is improved significantly and the correct singlet state is
now described (with an error in $\erw{\hat S^2}$ of $\sim\!10^{-4}$).
Similarly, while including $\hat P^{S^{2}}$ into the $D=1$ MMPS gave a qualitatively wrong PEC when using the R-AGP orbitals above,
after orbital optimization (dashed orange curve) we obtain the correct qualitative behavior.

As discussed in \autoref{sec:definitions} an alternative way to improve an MMPS other than increasing $D$ is to increase
the size of the sites. We find that using a $D=1$ multisite MMPS  (grouping two spatial orbitals into one site;  dark green curve) and optimizing the orbitals
  greatly improves the energies, compared to the GVB form, which makes a similar grouping but is more restricted (purple).\footnote{For this system in the minimal basis, GVB and the multisite MMPS sites both correspond to partitioning the wavefunction for the full problem into two fragments.}
  (Note that the GVB optimization included orbital optimization as well).

\subsection{\ce{O2} dissociation}
\label{sec:res_o2}

\begin{figure}[!tbp]
  \includegraphics[width=\columnwidth]{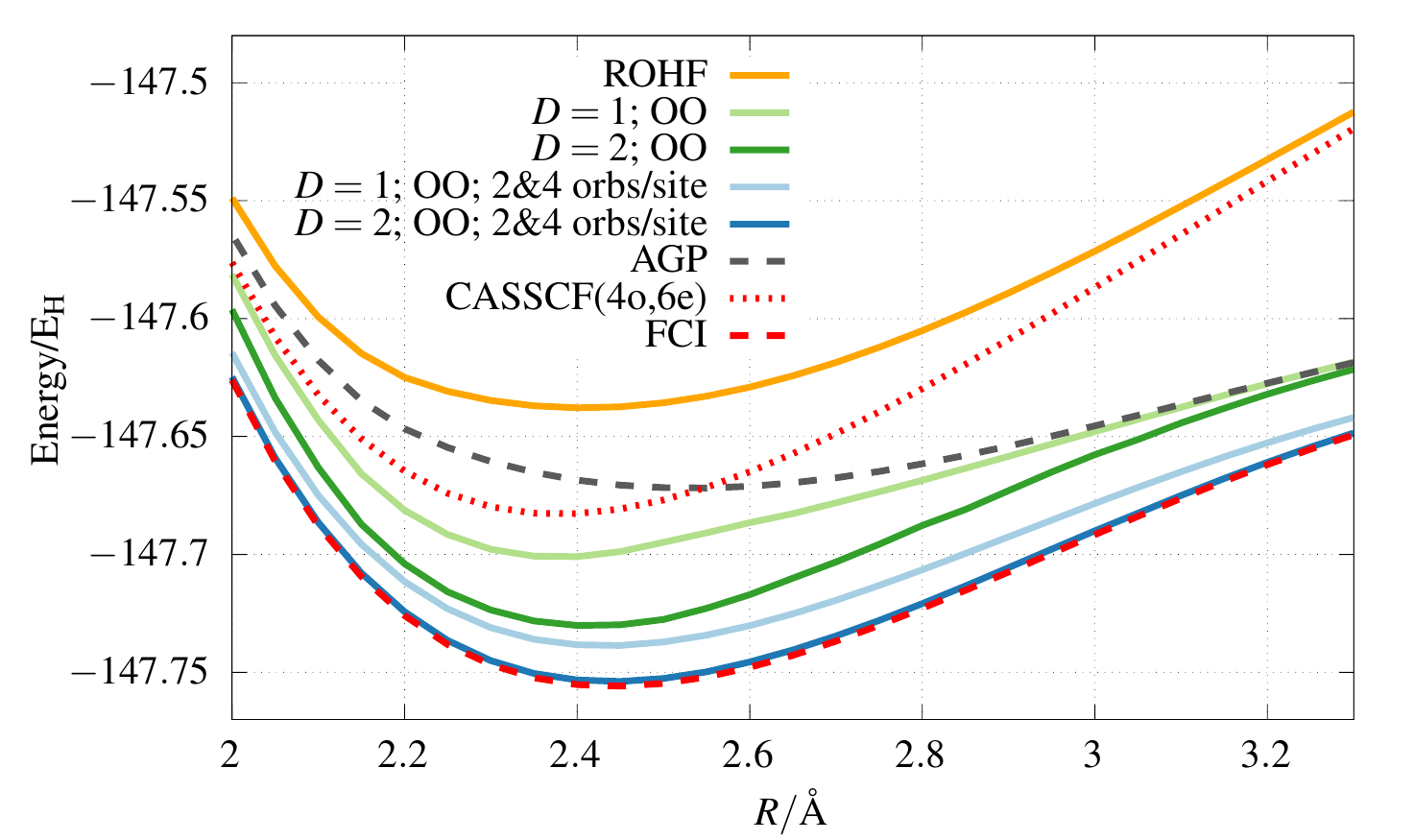}
  \caption{Potential energy curve of triplet \ce{O2} in STO-3G basis. Shown are results for restricted open-shell HF (ROHF, orange), {minimal matrix product state (MMPS) with orbital optimization (green), multisite MMPS (\new{blue}), restricted open-shell AGP (dashed gray), and complete active space-self consistent field, CASSCF, with a CAS consisting of four orbitals and six electrons (dotted red).
  \new{The $D=1$ orbitals are employed for the $D=2$ computations.}
  The results are compared to the full configuration interaction reference (FCI; dashed red curve). For the MMPS computation, $\hat P=\hat P^{S_z} \hat P^{N}$ is used as the projector.}
  }
  \label{fig:O2_pes}
\end{figure}

\ce{O2} is a prototypical open-shell multireference system.
The PEC of \ce{O2} in a STO-3G basis is shown in \autoref{fig:O2_pes}. For all bond distances shown, the FCI triplet state is the lowest state.
We see that the MMPS PECs (shown in green) are a significant improvement over the restricted open-shell AGP PEC (dashed gray curve).
The best \new{energies are} obtained by the multisite MMPS with one large site (\new{red} curves) consisting of four spatial orbitals
(to capture the minimal complete active space for triplet \ce{O2} which needs to contain four $2p$ orbitals)
and other large sites consisting of groups of two spatial orbitals.
\new{For $D=2$, this ansatz gives energies with a relative error of about $10^{-5}$, compared to FCI.}

Remarkably, all MMPSs, including the ones with $D=1$ and only 2 spin orbitals per site with either ordering, capture much more correlation energy than the minimal complete active space self-consistent field \new{calculation}, CASSCF(4o,6e), illustrating the compactness of the MMPS form.

\subsection{\ce{HF} dissociation}
\label{sec:res_fh}

\begin{figure*}[!tbp]
  \includegraphics[width=.6\textwidth]{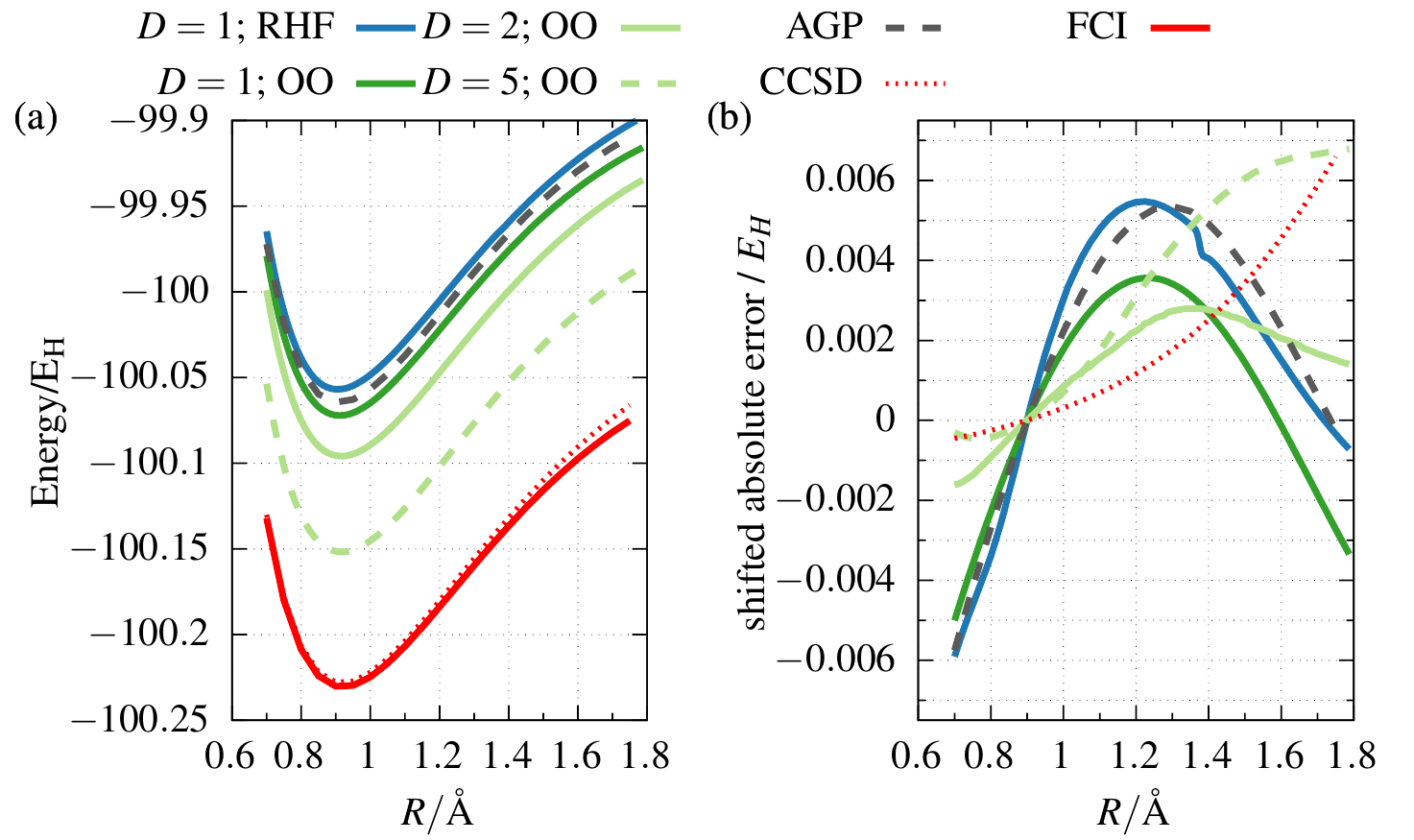}
    \caption{{Potential energy curve (PEC) (left panel) and \new{non-}parallelity errors across the PEC (right panel) for \ce{HF} with the cc-pVDZ basis. The
      \new{non-}parallelity errors are shifted absolute errors such that all PEC coincide at $R=\unit[0.9]{\textup{\AA}}$. The blue (green) curves denote the minimal matrix product state, MMPS, with restricted Hartree-Fock, RHF,  ($D=1$ optimized) orbitals, where $D$ is the bond dimension. 
    The dashed gray curve denotes the restricted antisymmetrized geminal product (AGP) curve. The dotted red curve corresponds to coupled cluster with singles and doubles (CCSD). 
    The results are compared to the full configuration interaction reference (FCI; dashed red curve). 
  For the MMPS computation, $\hat P=\hat P^{S_z}\hat P^{N}$ is used as the projector.}%
  }
  \label{fig:FH_pes}
\end{figure*}

To study the behavior of MMPS in non-minimal basis sets, we present results for the \ce{HF} PEC in the cc-pVDZ basis.\cite{ccpvdz}
\autoref{fig:FH_pes} shows the PEC and \new{non-}parallelity (shifted absolute) errors for this system. While for the bond distances shown,
the coupled cluster with singles and doubles (CCSD) method gives good results, the $D=1$ MMPS with just RHF orbitals (blue curve) actually
yields a similar \new{non-}parallelity error.
There is a small {``bump''} for the $D=1$ result with RHF orbitals at $R\sim\unit[1.37]{\textup{\AA}}$. This is near the Hartree-Fock
Coulson-Fischer point, but although the curve is bumpy we do not see a discontinuity
in the MMPS solution (i.e.~there is no sudden onset of symmetry breaking).
While an MMPS with $D=1$ with AGP orbitals (not shown) optimizes to give back the AGP wavefunction in this system (i.e.~all single
creation terms are zero), %
the MMPS with $D=1$ and optimized orbitals (dark green curve) results in an improved PEC. Orbital optimization also makes the ``bumps'' vanish.
Based on the optimized orbitals at $D=1$, increasing the bond dimension $D$ (pale green curves), gives additional substantial improvements both in the absolute ($D=2$ and $D=5$) and \new{non-}parallelity errors ($D=2$).

\section{Conclusions}
\label{sec:conclusions}

To summarize, we have explored a set of simple qualitative wavefunctions that we term minimal matrix product state\new{s} (MMPS). 
We define the MMPS to be an MPS with small bond dimension of $\bdim \sim 1$ combined with a projector onto the essential
symmetries of the problem, e.g.~particle, spin, and other symmetries. Already for $\bdim=1$, this framework includes
many other qualitative wavefunctions, such as symmetry broken and restored mean-field states, e.g.~projected Hartree-Fock and antisymmetrized geminal power states, and further extends them, e.g.~to beyond the seniority-zero sector in the case of the antisymmetrized geminal power.
Importantly, it does so while retaining the same computational scaling for energy evaluation and optimization as with such states.
This is because computations using the MMPS can use the density matrix renormalization group (DMRG) without
relying on the generalizations of Wick's theorem to incorporate symmetry projection. Similarly, the multisite version of the MMPS
extends generalized valence bond and strongly-orthogonal geminal wavefunctions and other related  ansätze beyond their product
state structure, via symmetry breaking and projection, as well as for $D>1$.

We examined the behaviour of MMPS in a number of prototypical systems, namely \ce{H4}, \ce{O2} and \ce{HF}.
The inclusion of the single creation operators is crucial to yield the observed improvements.
In all cases we found that the MMPS ansatz even with $D=1$ gives correct qualitative behavior of the potential energy landscape, often significantly improving on the aforementioned ansätze. 
We also noted that orbital optimization, an essential ingredient also of the other methods, significantly improves the MMPS wavefunction. In the cases where we increased $D$ but still kept it ``minimal'' ($\le 5$) we also observed a rapid improvement of the results.

We expect the MMPS ansatz to be useful in two main scenarios. 
First, MMPS could improve conventional DMRG calculations, which usually use large bond dimensions and
do not invoke projectors to restore symmetry, by serving as an initial guess state to improve optimization. An example of this can be found
in our previous work on spin-projected MPS,\cite{spinProjDMRG2017li,li2019electronic} which may be viewed through
the lens of this work as a type of MMPS.
Second, MMPS could serve as a method on its own for rapid exploration of the potential energy landscape\new{s} of molecular systems. 
This is especially useful for molecular dynamics simulations, where \new{there is a great need for fast electronic structure calculations.}
Possible extensions to treat dynamical correlation~\cite{pDMRG2018guo,pDMRGstoch2018guo,nevpt2Dmrg2017freitag,nevpt2Dmrg2016chan,dftDmrg2015reiher,gvbRPA2019pernal} and excited states~\cite{lrDMRG2014nakatani,exStateGeomOptDMRG2015hu,dorando_2007,baiardi_2019} are possible as well.
Further, the methodology can straightforwardly be transferred to related domains, most importantly in applications to quantum dynamics.\cite{mctdhXrev2020lode,mctdhSQRn2020weike,ttn2019larsson,cp2014leclerc}

\if\USEACHEMSO1
\begin{acknowledgement}
\else
\acknowledgements
\fi
This work was supported by the US NSF via grant no.~CHE-1665333.
HRL acknowledges support from the German Research Foundation (DFG) via grant LA 4442/1-1.
CAJH acknowledges support from a generous start-up package from
Wesleyan University.
\if\USEACHEMSO1
\end{acknowledgement}
\fi

\if\USEACHEMSO1
\clearpage
For Table of Contents Only\\
{\centering
\includegraphics{pics/cover}}
\fi

\end{document}